\documentstyle[12pt]{article}
\newcommand{\R}{\rm I \mkern -3mu R}
\begin{document}
\thispagestyle{empty}
\begin{flushright}
IFA-FT-391-1993, December
\end{flushright}
\bigskip\bigskip\begin{center}
{\bf \Huge{THE PROJECTIVE UNITARY IRREDUCIBLE REPRESENTATIONS\\}}\vskip1mm
{\bf \Huge{OF THE GALILEI GROUP IN 1+2 DIMENSIONS}}
\end{center}
\vskip 1.0truecm
\centerline{\bf
D. R. Grigore\footnote{e-mail: grigore@roifa.bitnet}}
\vskip5mm
\centerline{Dept. Theor. Phys., Inst. Atomic Phys.,}
\centerline{Bucharest-M\u agurele, P. O. Box MG 6, ROM\^ANIA}
\vskip 2cm
\bigskip \nopagebreak \begin{abstract}
\noindent
We give an elementary analysis of the multiplicator group of the
Galilei group in 1+2 dimensions
$G^{\uparrow}_{+}$.
For a non-trivial multiplicator we give a list of all the
corresponding projective unitary irreducible representations of
$G^{\uparrow}_{+}$.
\end{abstract}
\vskip 2cm
\vskip 3truecm
Shortened title: Galilei group in 1+2 dimensions

\newpage\setcounter{page}1

\section{Introduction}

Recently we have determined a complete list of all projective
unitary irreducible representations of the Poincar\'e group in
$1+2$ dimensions \cite{G}. In this paper we intend to provide a
similar analysis for the Galilei group.

The main technical obstacle seems to be the rather complicated
structure of the multiplicator group (see \cite{MS}, Appendix A). In
Section 2 we give an elementary analysis of the second
cohomology group of the universal covering group
$
\widetilde{G^{\uparrow}_{+}}
$
of the Galilei group
$G^{\uparrow}_{+}$
in $1+2$ dimensions.

In Section 3 we construct for every non-trivial multiplicator a
certain  extension of
$
\widetilde{G^{\uparrow}_{+}}
$
which exhibits a semi-direct product
group structure. Then we are able to apply Mackey induced
representations method to determine the desired representations.
In this paper only non-trivial multiplicators are considered.
The case of true representation (i.e. trivial multiplicators) is
elementary to analyse and raises no problems \cite{G}.

\section{The Galilei group in 1+2 dimensions}

\subsection*{A. Notations}

By definition the ortochronous Galilei group in $1+2$ dimensions
$G^{\uparrow}$
is set-theoretically
$O(2) \times \R^{2} \times \R \times \R^{2}$
with the composition law:
$$
(R_{1},{\bf v}_{1},\eta_{1},{\bf a}_{1}) \cdot
(R_{2},{\bf v}_{2},\eta_{2},{\bf a}_{2}) =
(R_{1}R_{2},R_{1}{\bf v}_{2}+{\bf v}_{1},\eta_{1}+\eta_{2},
R_{1}{\bf a}_{2}+{\bf a}_{1}+\eta_{2}{\bf v}_{1}).
\eqno(2.1)$$

We organize
$\R^{2}$
as column vectors,
$O(2)$
as the
$2 \times 2$
real orthogonal matrices and we use consistently matrix
notations. This group acts naturally on
$\R \times \R^{2}$:
$$
(R,{\bf v},\eta,{\bf a}) \cdot(T,{\bf X}) =
(T+\eta,R{\bf X}+T{\bf v}+{\bf a}).
\eqno(2.2)$$

We will also consider the proper orthochronous Galilei group
$G^{\uparrow}_{+}$
defined as:
$$
G^{\uparrow}_{+} \equiv \{(R,{\bf v},\eta,{\bf a}) \vert det(R)
= 1 \}
\eqno(2.3)$$
and the universal covering group
$\widetilde{G^{\uparrow}_{+}}$ of
$G^{\uparrow}_{+}$.
We can take
$
\widetilde{G^{\uparrow}_{+}} =
\R \times \R^{2} \times \R \times \R^{2}
$
with the composition law:
$$
(x_{1},{\bf v}_{1},\eta_{1},{\bf a}_{1}) \cdot
(x_{2},{\bf v}_{2},\eta_{2},{\bf a}_{2}) =
$$
$$
(x_{1}+x_{2},{\bf v}_{1}+R(x_{1}){\bf v}_{2},\eta_{1}+\eta_{2},
{\bf a}_{1}+R(x_{1}){\bf a}_{2}+\eta_{2}{\bf v}_{1}),
\eqno(2.4)$$
where
$$
R(x) \equiv \left( \matrix{cos(x) & sin(x) \cr -sin(x) &
cos(x)}\right)
\eqno(2.5)$$

The covering homomorphism
$\delta:\widetilde{G^{\uparrow}_{+}} \rightarrow G^{\uparrow}_{+}$
is:
$$
\delta(x,{\bf v},\eta,{\bf a}) = (R(x),{\bf v},\eta,{\bf a}).
\eqno(2.6)$$

Finally we describe the Lie algebra of
$
\widetilde{G^{\uparrow}_{+}}
$,
$
Lie(\widetilde{G^{\uparrow}_{+}}) \simeq Lie(G^{\uparrow}_{+})
$
using the fact that
$
G^{\uparrow}_{+}
$
can be organized as a matrix group. Indeed one has the group
isomorphism:
$$
G^{\uparrow}_{+} \ni (R,{\bf v},\eta,{\bf a}) \leftrightarrow
\left( \matrix{ R & {\bf v} & {\bf a} \cr 0 & 1 & \eta \cr 0 & 0
& 1} \right) \in M_{\R}(4,4).
\eqno(2.7)$$

Then
$
Lie(G^{\uparrow}_{+})
$
can be identified with the linear space of
$
4 \times 4
$
real matrices of the form:
$$
(\alpha,{\bf u},t,{\bf x}) \equiv \left( \matrix{\alpha A & {\bf
u} & {\bf x} \cr 0 & 0 & t \cr 0 & 0 & 0}\right).
\eqno(2.8)$$

Here
$
{\bf u}, {\bf x} \in \R^{2}, t, \alpha \in \R,
A \equiv \left( \matrix{ 0 & 1 \cr -1 & 0}\right),
$
and the exponential map is the usual matrix exponential. One can
easily obtains the Lie bracket as:
$$
[(\alpha_{1},{\bf u}_{1},t_{1},{\bf x}_{1}),
(\alpha_{2},{\bf u}_{2},t_{2},{\bf x}_{2})] =$$
$$(0,A(\alpha_{1} {\bf u}_{2}-\alpha_{2} {\bf u}_{1}),0,
A(\alpha_{1} {\bf x}_{2}-\alpha_{2} {\bf x}_{1})+
t_{2} {\bf u}_{1}-t_{1} {\bf u}_{2}).
\eqno(2.9)$$

\subsection*{B. Computation of
$
H^{2}(Lie(\widetilde{G^{\uparrow}_{+}}),\R)
$}

As it is well known, to classify all multipliers of a Lie group
$G$ one has to compute first the second cohomology group of
$Lie(G)$
with real coefficients \cite{V}. As we have said in the
Introduction, we provide here an elementary derivation of this
group. If
$
\xi \in Z^{2}(Lie(\widetilde{G^{\uparrow}_{+}}),\R)
$
the cocycle equation writes as:
$$
\xi([(\alpha_{1},{\bf u}_{1},t_{1},{\bf x}_{1}),
(\alpha_{2},{\bf u}_{2},t_{2},{\bf x}_{2})],
(\alpha_{3},{\bf u}_{3},t_{3},{\bf x}_{3})) + circular~~
permutations = 0.
\eqno(2.10)$$

Here
$
\xi: Lie(\widetilde{G^{\uparrow}_{+}}) \times
Lie(\widetilde{G^{\uparrow}_{+}}) \rightarrow \R
$
is, by definition, bilinear and antisymmetric. Using (2.9), we
can explicitate (2.10):
$$
\xi((0,A(\alpha_{1} {\bf u}_{2}-\alpha_{2} {\bf u}_{1}),0,
A(\alpha_{1} {\bf x}_{2}-\alpha_{2} {\bf x}_{1})+
t_{2} {\bf u}_{1}-t_{1} {\bf u}_{2}),
(\alpha_{3},{\bf u}_{3},t_{3},{\bf x}_{3})) +
$$
$$
+ circular~~permutations = 0.
\eqno(2.11)$$

We have to consider some distinct cases of this equation:

(i)~~$t_{i} = 0, {\bf x}_{i} = {\bf 0}~~( i = 1,2,3)$

One obtains:
$$
\xi((0,A(\alpha_{1} {\bf u}_{2}-\alpha_{2} {\bf u}_{1}),0,
{\bf 0})) + circular~~permutations = 0.
\eqno(2.12)$$

{}From bilinearity we have:
$$
\xi((0,{\bf u},0,{\bf 0}),(0,{\bf u'},0,{\bf 0})) =
{\bf u}^{t} C {\bf u'}
$$
where $C$ is a
$
2 \times 2
$
real matrix. From antisymmetry we find
$
C^{t} = - C
$
so necessarely
$
C = \frac{1}{2} F A~~(F \in \R).
$
So we have:
$$
\xi((0,{\bf u},0,{\bf 0}),(0,{\bf u'},0,{\bf 0}) =
\frac{1}{2}~F <{\bf u},{\bf u'}>
\eqno(2.13)$$
where
$<\cdot,\cdot>$
is the sesquilinear form on
$
\R^{2}
$
given by:
$$
<{\bf u},{\bf v}> \equiv
{\bf u}^{t} A {\bf v}
\eqno(2.14)$$

It is easy to see that (2.12) becomes an identity.

(ii) $\alpha_{i} = 0,~{\bf u}_{i} = {\bf 0}~~(i = 1,2,3)$

Equation (2.11) becomes an identity.

(iii) $t_{1} = 0,~{\bf x}_{1} = {\bf 0},~\alpha_{i} = 0,~{\bf
u}_{i} = {\bf 0}~~(i = 1,2)$

One easily obtains from (2.11):
$$
\xi((0,{\bf 0},t,{\bf x}),(0,{\bf 0},t,{\bf x'})) = 0.
\eqno(2.15)$$

(iv) $t_{i} = 0,~{\bf x}_{i} = {\bf 0}~(i = 1,2),~
\alpha_{3} = 0,~{\bf u}_{3} = {\bf 0}$

Equation (2.11) becomes:
$$
\xi((0,A(\alpha_{1} {\bf u}_{2}-\alpha_{2} {\bf u}_{1}),0,
{\bf 0}), (0,{\bf 0},t_{3},{\bf x}_{3})) +$$
$$\xi((0,{\bf 0},0,\alpha_{2} A {\bf x}_{3}+t_{3} {\bf
u}_{2}),(\alpha_{1},{\bf u_{1}},0,{\bf 0})) +$$
$$\xi((0,{\bf 0},0,-\alpha_{1} A {\bf x}_{3}-t_{3} {\bf u}_{1}),
(\alpha_{2},{\bf u_{2}},0,{\bf 0})) =
0.\eqno(2.16)$$

{}From bilinearity one has:
$$
\xi((0,{\bf u},0,{\bf 0}),(0,{\bf 0},0,{\bf x}) =
{\bf x}^{t} D {\bf u}
\eqno(2.17)$$
where
$D$
is a
$
2 \times 2
$
real matrix.

If we take in (2.16)
$
\alpha_{2} = 0,~t_{3} = 0
$
and we insert the expression (2.17) we get
$$
[D,A] = 0 \Leftrightarrow D = -\tau \times id + c A
{}~~(\tau, c \in \R)
$$
so, (2.17) takes the form:
$$
\xi((0,{\bf u},0,{\bf 0}),(0,{\bf 0},0,{\bf x}) =
\tau {\bf x} \cdot {\bf u} + c <{\bf x},{\bf u}>.
\eqno(2.18)$$

If we insert (2.18) into (2.16) we get
$
c = 0
$
and:
$$
\xi((0,A{\bf u},0,{\bf 0}),(0,{\bf 0},1,{\bf 0})) +
\xi((0,{\bf 0},0,{\bf u}),(1,{\bf 0},0,{\bf 0})) = 0
$$

Because of linearity we have:
$$
\xi((0,{\bf 0},1,{\bf 0}),(0,{\bf u},0,{\bf 0})) =
{\bf P}\cdot {\bf u}~~~({\bf P} \in \R^{2})
\eqno(2.19)$$
and the preceeding relation gives:
$$
\xi((0,{\bf 0},0,{\bf x}),(1,{\bf 0},0,{\bf 0})) = <{\bf P},{\bf
x}>
\eqno(2.20)$$

Finally (2.18) reduces to:
$$
\xi((0,{\bf u},0,{\bf 0}),(0,{\bf 0},0,{\bf x}) =
\tau {\bf x} \cdot {\bf u}.
\eqno(2.21)$$

We denote:
$$
S \equiv \xi((1,{\bf 0},0,{\bf 0}),(0,{\bf 0},1,{\bf 0}))
\eqno(2.22)$$
and use linearity to obtain:
$$
\xi((1,{\bf 0},0,{\bf 0}),(0,{\bf u},0,{\bf 0}) = <{\bf G},{\bf
u}>~~~({\bf G} \in \R^{2}).
\eqno(2.23)$$

Finally we can reconstruct the cocyle
$\xi$
from (2.13), (2.15), (2.17) and (2.19)-(2.23):
$$
\xi(\alpha_{1},{\bf u}_{1},t_{1},{\bf x}_{1}),
(\alpha_{2},{\bf u}_{2},t_{2},{\bf x}_{2})) =
<{\bf G},\alpha_{1} {\bf u}_{2}-\alpha_{2} {\bf u}_{1}> +
\frac{1}{2} F <{\bf u}_{1},{\bf u}_{2}> +$$
$$S (\alpha_{1} t_{2} - \alpha_{2}
t_{1}) - <{\bf P},A( \alpha_{1} {\bf x}_{2}-\alpha_{2} {\bf x}_{1})
+ t_{2} {\bf u}_{1}-t_{1} {\bf u}_{2}> + \tau ({\bf x}_{1}\cdot
{\bf u}_{2} - {\bf x}_{1}\cdot {\bf u}_{2})
\eqno(2.24)$$

If we take the generic element of
$
C^{1}(Lie(\widetilde{G^{\uparrow}_{+}}),\R)
\simeq (Lie(\widetilde{G^{\uparrow}_{+}})^{*}
$
to be
$
(\beta,{\bf G},E,{\bf P})
$
defined by:
$$
<(\beta,{\bf G},E,{\bf P}),(\alpha,{\bf u},t,{\bf x})> =
-\beta \alpha - {\bf G}\cdot {\bf u} -Et + {\bf P}\cdot {\bf x}
\eqno(2.25)$$
then it is elementary to see that (2.24) rewrites as:
$$
\xi = \tau \xi_{0} + F \xi_{1} + S \xi_{2} + \partial (0,{\bf
G},0,{\bf P})
\eqno(2.26)$$
where
$
\xi_{0}, \xi_{1}, \xi_{2} \in
Z^{2}(Lie(\widetilde{G^{\uparrow}_{+}}),\R)
$
are:
$$
\xi_{0}((\alpha_{1},{\bf u}_{1},t_{1},{\bf x}_{1}),
(\alpha_{2},{\bf u}_{2},t_{2},{\bf x}_{2}) = {\bf x}_{1}\cdot
{\bf u}_{2} - {\bf x}_{1}\cdot {\bf u}_{2}
\eqno(2.27)$$
$$
\xi_{0}((\alpha_{1},{\bf u}_{1},t_{1},{\bf x}_{1}),
(\alpha_{2},{\bf u}_{2},t_{2},{\bf x}_{2}) =
\frac{1}{2}  <{\bf u}_{1},{\bf u}_{2}>
\eqno(2.28)$$
$$
\xi_{0}((\alpha_{1},{\bf u}_{1},t_{1},{\bf x}_{1}),
(\alpha_{2},{\bf u}_{2},t_{2},{\bf x}_{2}) =
\alpha_{1} t_{2} - \alpha_{2} t_{1}
\eqno(2.29)$$

So every cocycle
$
\xi \in Z^{2}(Lie(\widetilde{G^{\uparrow}_{+}}),\R)
$
is cohomologous with a cocyle of the form
$
\tau \xi_{0} + F \xi_{1} + S \xi_{2}.
$

It is easy to establish now that this cocycle is not a
coboundary. So we can summarize the preceeding discussion as:

{\bf Proposition 1:}
$
H^{2}(Lie(\widetilde{G^{\uparrow}_{+}}),\R)
$
is a three dimensional real space. In every cohomology class
there exists exactly one cocyle of the type
$
\tau \xi_{0} + F \xi_{1} + S \xi_{2}
$.

\subsection*{C. Computation of
$
H^{2}(\widetilde{G^{\uparrow}_{+}},\R)
$}

As in \cite{V} one uses the fact that for $G$ a connected and
simply connected Lie group,
$
H^{2}(G,\R)
$
is isomorphic to
$
H^{2}(Lie(G),\R).
$
So, to determine
$
H^{2}(G,\R)
$
one should determine for
$
\xi_{0}, \xi_{1}, \xi_{2}
$
some corresponding group cocyles. One easily determines (see
\cite{MS}):
$$
\omega_{0}(g,g') = 1/2~[{\bf a}\cdot R(x) {\bf v'} - {\bf
v}\cdot R(x) {\bf a'} + t' {\bf v}\cdot R(x) {\bf v'}]
\eqno(2.30)$$
$$
\omega_{1}(g,g') = 1/2 < {\bf v}, R(x) {\bf v'} >
\eqno(2.31)$$
$$
\omega_{2} = \eta x'
\eqno(2.32)$$
where:
$
g = (x,{\bf v},\eta,{\bf a}),~g' = (x',{\bf v'},\eta',{\bf a'})
$

So we have:

{\bf Corollary 1:}
$
H^{2}(\widetilde{G^{\uparrow}_{+}},\R)
$
is a three dimensional real linear space. In every cohomology
class there exists exactly one cocycle of the type
$
\tau \omega_{0} + F \omega_{1} +S \omega_{2}.
$

Finally, applying again Thm. 7.37 of \cite{V} we have:

{\bf Corollary 2:}
Every multiplier of
$\widetilde{G^{\uparrow}_{+}}$
is equivalent to a multiplier of the form
$
m_{\tau} m_{F} m_{S}
$
where:
$$
m_{\tau}(g,g') \equiv exp\left\{\frac{i}{2} [{\bf a}\cdot R(x) {\bf v'}
- {\bf v}\cdot R(x) {\bf a'} + t' {\bf v}\cdot R(x) {\bf v'}]\right\}
\eqno(2.33)$$
$$
m_{F}(g,g') \equiv exp\left\{\frac{i}{2} < {\bf v}, R(x) {\bf v'} >\right\}
\eqno(2.34)$$
$$
m_{S}(g,g') \equiv e^{i \eta x'}
\eqno(2.35)$$

{\bf Remark 1:} In 1+3 dimensions it is well known that only a
multiplier of type (2.33) survives. So,
$m_{F}$
and
$m_{S}$
are characteristic to 1+2 dimensions. According to the general
theory of the projective unitary representations, one has to
consider the most general multiplier, which in this case is
$
m_{\tau} m_{F} m_{S}.
$
In \cite{G} only the multiplier
$
m_{\tau}
$
was considered, so the statement of Thm. 2 is rather careless.

\section{ The projective unitary irreducible representations of
the Galilei group in 1+2 dimensions}

\subsection*{A. The method}

As anticipated, we classify here the unitary irreducible
$
m_{\tau} m_{F} m_{S}-
$
representations of
$G^{\uparrow}_{+}$.
We try to mimick the method form \cite{V} which consists of two
steps:

a) One first applies a simple result, namely a generalization of
Thm. 7.16 from \cite{V}.

{\bf Proposition 2:}
Let
$
m_{0},...,m_{p}
$
be multipliers for the group $G$ and let
$
U~(g \mapsto U_{g})
$
be an
$
m_{0}\cdots m_{p}-
$
representation of $G$ in a (separable) Hilbert
${\cal H}$
space over the
complex numbers. Let
$
G_{m_{0},...,m_{p}} \equiv G \times \underbrace{{\bf T} \times \cdots
\times {\bf T}}_{(p+1)-times}
$
(where ${\bf T}$ is the set of complex numbers of modulus 1 considered
as a multiplicative group) with the composition law:
$$
(g;\zeta_{0},...,\zeta_{p})\cdot (g';\zeta'_{0},...,\zeta'_{p}) =
(gg';\zeta_{0}\zeta'_{0}m_{0}(g,g'),...,\zeta_{p}\zeta'_{p}m_{p}(g,g'))
\eqno(3.1)$$

Then
$
G_{m_{0},...,m_{p}}
$
is a group, and:
$$
(g;\zeta_{0},...,\zeta_{p}) \mapsto
V_{g;\zeta_{0},...,\zeta_{p}} \equiv \zeta_{0}^{-1}\cdots
\zeta_{p}^{-1} U_{g}
\eqno(3.2)$$
is a representation of
$
G_{m_{0},...,m_{p}}
$
in
${\cal H}$
such that:
$$
V_{e;\zeta_{0},...,\zeta_{p}} = \zeta_{0}^{-1}\cdots
\zeta_{p}^{-1} \times id
\eqno(3.3)$$

Conversely, if $V$ is a representation of
$
G_{m_{0},...,m_{p}}
$
in
${\cal H}$
such that (3.3) is verified, then if we write:
$$
U_{g} \equiv V_{g;\underbrace{1,...,1}_{(p+1)-times}}
\eqno(3.4)$$
$
g \mapsto U_{g}
$
is an
$
m_{0}\cdots m_{p}-
$
representation $G$ in
${\cal H}$
and the connection (3.2) between $U$ and $V$ is true.

The proof is elementary and Thm. 7.16 of \cite{V} is the
particular case
$p = 0$.

Is is clear that we will apply Prop. 2 with
$p = 2$,
and
$
m_{0} = m_{\tau},~m_{1} = m_{F},~m_{2} = m_{S}.
$
We will denote the the corresponding group
$
G_{m_{0}m_{1}m_{2}}
$
by
$
G^{\tau,F,S}.
$

b) We remember that in \cite{V}, where one has to study only
$
G^{\tau,0,0}.
$
(see ch. IX, section 8), a semi-direct product group structure
is exhibited, so one can apply the induction procedure. The same
procedure works for
$
G^{\tau,F,S}.
$
Indeed one has an group isomorphism
$
G^{\tau,F,S} \simeq H^{F} \times_{t^{\tau,S}} A^{\tau,S}
$
where:

- Let us define the group
$
H \equiv \R \times \R^{2}
$
with the composition law:
$$
(x,{\bf v}) \cdot (x',{\bf v'}) = (x+x',{\bf v}+R(x){\bf v'})
\eqno(3.5);$$
then
$
H^{F} \equiv H \times {\bf T}
$
with the composition law:
$$
(h_{1};\zeta_{1}) \cdot (h_{2};\zeta_{2}) =
(h_{1}h_{2};\zeta_{1} \zeta'_{1} m_{F}(h,h'))
\eqno(3.6)$$

The notation
$
m_{F}(h,h')
$
makes sense because in the right hand side of (2.34) only the
variables
$x$ and ${\bf v}$
appear.

- $A^{\tau,S} \equiv \R \times \R^{2} \times {\bf T} \times {\bf T}$
with the composition law:
$$
(\eta,{\bf a};\zeta_{0},\zeta_{2}) \cdot
(\eta',{\bf a'};\zeta'_{0},\zeta'_{2}) = (\eta+\eta',{\bf a}+{\bf a'};
\zeta_{0}\zeta'_{0},\zeta_{2}\zeta'_{2})
\eqno(3.7)$$

- $t^{\tau,S}: H^{F} \rightarrow Aut \left( A^{\tau,S} \right)$
is given by:
$$
t^{\tau,S}_{h;\zeta_{1}}(a;\zeta_{0},\zeta_{2}) =
\left( t_{h}(a);\zeta_{0}exp\left\{-\frac{i\tau}{2} [ 2{\bf v}\cdot R(x){\bf
a}+\eta{\bf v}^{2}] \right\},\zeta_{2} e^{-iS\eta x} \right)
\eqno(3.8)$$

Here
$a \equiv (\eta,{\bf a})$
and
$$
t_{h}(\eta,{\bf a}) = (\eta,R(x){\bf a}+\eta{\bf v})
\eqno(3.9)$$

The semi-direct product structure follows from the homomorphism
property of
$t^{\tau,S}$.
Then the isomorphism is:
$$
H^{F} \times_{t^{\tau,S}} A^{\tau,S} \ni \{(x,{\bf v};\zeta_{1}),
(\eta,{\bf a};\zeta_{0},\zeta_{2})\} \leftrightarrow$$
$$\left(x,{\bf v},\eta,{\bf a};\zeta_{0} exp\left( \frac{i\tau}{2}
{\bf a}\cdot {\bf v}\right),\zeta_{1},\zeta_{2} e^{iS\eta x}\right)
\in G^{\tau,F,S}
\eqno(3.10)$$
We note that
$A^{\tau,S}$
is Abelian, so we will be able to apply Mackey induction procedure.

{\bf Remark 2:} One may wonder why we did not apply Prop. 2 with
$p = 0$
and
$
m_{0} = m_{\tau} m_{F} m_{S}.
$
The reason is that we did not succeed to find a convenient
semi-direct product structure for
$G_{m_{0}}$
in this case.

Taking into account (3.10) the representations of
$
G^{\tau,F,S}
$
follow from representations of
$
H^{F} \times_{t^{\tau,S}} A^{\tau,S}
$
according to:
$$
W_{(x,{\bf v},\eta,{\bf a};\zeta_{0},\zeta_{1},\zeta_{2})} =
{\cal W}_{\left(x,{\bf v},\eta,{\bf a};\zeta_{0}exp\left\{-\frac{i\tau}{2}
{\bf a}\cdot {\bf v}\right\},\zeta_{1},\zeta_{2}e^{-iS\eta x}\right))}
\eqno(3.11)$$

According to Prop. 2 we are looking for unitary irreducible
representations of
$
G^{\tau,F,S}
$
verifying:
$$
W_{e;\zeta_{0},\zeta_{1},\zeta_{2}} = \zeta_{0}^{-1}
\zeta_{1}^{-1} \zeta_{2}^{-1} \times id
\eqno(3.12)$$

Now we will follow the method of induced representations as
presented in section II C of \cite{G}.

\subsection*{B. Computation of the orbits}

It is clear that every character of
$A^{\tau,S}$
has the form:
$$
\chi_{p_{0},{\bf p};n_{0},n_{2}}(\eta,{\bf
a};\zeta_{0},\zeta_{2}) = \zeta_{0}^{n_{0}} \zeta_{2}^{n_{2}}
exp\{i(\eta p_{0}+{\bf a}\cdot {\bf p})\}
\eqno(3.13)$$
where
$
p_{0} \in \R,~{\bf p} \in \R^{2}
$
and
$n_{0},n_{2} \in {\bf Z}.$
So,
$
\widehat{A^{\tau,S}} \equiv \R \times \R^{2} \times {\bf Z}
\times {\bf Z}
$
with the generic element denoted by
$
[p_{0},{\bf p};n_{0},n_{2}].
$
One easily computes the dual action of
$t_{h}^{\tau,S},$
namely:
$$
(x,{\bf v};\zeta_{1})\cdot [p_{0},{\bf p};n_{0},n_{2}] = $$
$$\left[p_{0}-{\bf v}\cdot R(x){\bf p}-1/2 n_{0}\tau{\bf
v}^{2}+Sn_{2}x,R(x){\bf p}+n_{0}\tau{\bf v};n_{0},n_{2}\right]
\eqno(3.14)$$

The classification of the orbits of this action is elementary.
We distinguish three cases which must be studied separatedly:

a) $\tau \not= 0,~S = 0$

The orbits are:

$$
Z^{1}_{n_{2},p_{0}} \equiv \{ [p_{0},{\bf 0};0,n_{2}] \};~p_{0} \in
\R,~n_{2} \in {\bf Z}
$$
$$
Z^{2}_{n_{2},\rho} \equiv \{ [p_{0},{\bf p};0,n_{2}] \vert
{\bf p}^{2} = r^{2},p_{0} \in \R \}; r \in \R_{+},~n_{2} \in {\bf Z}
$$
$$
Z^{3}_{n_{0},n_{2},\rho} \equiv \{ [p_{0},{\bf p};n_{0},n_{2}] \vert
{\bf p}^{2}+2n_{0}\tau p_{0} = \rho \};~\rho \in \R,~
n_{0} \in {\bf Z}^{*}, n_{2} \in {\bf Z}
$$

b) $\tau = 0,~S \not= 0$

$$
Z^{4}_{n_{0},p_{0}} \equiv \{ [p_{0},{\bf 0};n_{0},0] \};~p_{0} \in
\R,~n_{0} \in {\bf Z}
$$
$$
Z^{5}_{n_{0},n_{2}} \equiv \{ [p_{0},{\bf 0};n_{0},n_{2}] \vert
p_{0} \in \R \},~n_{0} \in {\bf Z},~n_{2} \in {\bf Z}^{*}
$$
$$
Z^{6}_{n_{0},n_{2},\rho} \equiv \{ [p_{0},{\bf p};n_{0},n_{2}] \vert
{\bf p}^{2} = r^{2},p_{0} \in \R \}; r \in \R_{+},
{}~n_{0},n_{2} \in {\bf Z}
$$

c) $\tau \not= 0,~S \not= 0$

$$
Z^{7}_{p_{0}} \equiv \{ [p_{0},{\bf 0};0,0] \};~p_{0} \in \R
$$
$$
Z^{8}_{n_{2}} \equiv \{ [p_{0},{\bf 0};0,n_{2}] \vert
p_{0} \in \R \},n_{2} \in {\bf Z}^{*}
$$
$$
Z^{9}_{n_{0},\rho} \equiv \{ [p_{0},{\bf p};n_{0},0] \vert
{\bf p}^{2}+2n_{0}\tau p_{0} = \rho \};~\rho \in \R,~
n_{0} \in {\bf Z}^{*}
$$
$$
Z^{10}_{n_{0},n_{2}} \equiv \{ [p_{0},{\bf p};n_{0},n_{2}] \vert
{\bf p} \in \R^{2}, p_{0} \in \R \}; ~n_{0},n_{2} \in {\bf Z}^{*}
$$

It is not hard to see that condition (3.12) cannot be fulfilled
by the induced representations corresponding to the orbits:
$
Z^{1}, Z^{2}, Z^{4}, Z^{7}, Z^{8}
$
and
$Z^{9}.$
So, we have to analyse only the cases
$
Z^{3}, Z^{5}, Z^{6}
$
and
$Z^{10}$.
We analyse one by one these four cases.

\subsection*{C. The representations}

a) $\tau \not= 0,~S = 0$

In this case,
$
Z = Z^{3}_{n_{0},n_{2},\rho}.
$
A refernce point is
$
\left[\frac{\rho}{2n_{0}\tau},{\bf 0};n_{0},n_{2}\right]
$
and we have:
$$
H^{F}_{\left[\frac{\rho}{2n_{0}\tau},{\bf 0};n_{0},n_{2}\right]} =
\{(x,{\bf 0};\zeta_{1} \vert x \in \R,~\zeta_{1} \in {\bf T} \}
\simeq \R \times {\bf T}
$$

The unitary irreducible representations of this Abelian subgroup
are of the form
$
\pi^{(s,n_{1})}~(s \in \R,~n_{1} \in {\bf Z});
$
they are one-dimensional, are acting in
${\bf C}$
as follows:
$$
\pi^{(s,n_{1})}(x,{\bf 0};\zeta_{1}) = e^{isx}~\zeta_{1}^{n_{1}}
\eqno(3.15)$$

In this case one can find explicitely a corresponding cocycle
$\phi^{\pi}$,
namely:
$$
\phi^{(s,n_{1})}((x,{\bf v};\zeta_{1}),[p_{0},{\bf
p};n_{0},n_{2}]) = e^{isx}~\zeta_{1}^{n_{1}}~exp\left\{
\frac{in_{1}F}{2n_{0}\tau} <{\bf v},R(x){\bf p}>\right\}
\eqno(3.16)$$

As in \cite{V} we identify
$
Z^{3} \simeq \R^{2}
$
according to:
$$
[(\rho-{\bf p}^{2})/2n_{0}\tau,{\bf p};n_{0},n_{2}]
\leftrightarrow {\bf p}
$$
and we consider the strictly invariant Lebesgue measure
$
d{\bf p}
$
on
$\R^{2}$.

Applying (3.11), the corresponding induced representation is
acting in
$
{\cal H} = L^{2}(\R^{2},d{\bf p})
$
according to:
$$
\left(W_{(x,{\bf v},\eta,{\bf
a};\zeta_{0},\zeta_{1},\zeta_{2})}f\right) ({\bf p}) =$$
$$\zeta_{0}^{n_{0}}\zeta_{1}^{n_{1}}\zeta_{2}^{n_{2}}
exp\left\{i\left[-\frac{n_{0}\tau}{2} {\bf a}\cdot{\bf v}+\eta
\frac{\rho-{\bf p}^{2}}{2n_{0}\tau}+ {\bf a}\cdot{\bf p}+sx+
\frac{n_{1}F}{2n_{0}\tau} <{\bf v},{\bf p}>\right]\right\}$$
$$f(R(x)^{-1}({\bf p}-n_{0}\tau{\bf v})
$$

The condition (3.12) imposes
$
n_{0} = n_{1} = n_{2} = -1.
$
The factor
$
exp\left(-\frac{i\eta\rho}{2\tau}\right)
$
can be dropped because we are loking for projective
representations. We get the projective representations
$
V^{\tau,s}~~~(\tau \in \R^{*}, s \in \R)
$
acting in
$
L^{2}(\R^{2},d{\bf p})
$
as follows:
$$
\left(V^{\tau,s}_{x,{\bf v},\eta,{\bf a}}f\right)({\bf p}) =$$
$$exp\left\{i\left( sx+{\bf a}\cdot{\bf p}+\frac{\eta{\bf p}^{2}}{2\tau}
+\frac{\tau}{2}{\bf a}\cdot{\bf v}+\frac{F}{2\tau}<{\bf v},{\bf p}>
\right)\right\} f(R(x)^{-1}({\bf p}+\tau{\bf v})).
\eqno(3.17)$$

b) $\tau = 0,~S \not= 0$

b1) $Z = Z^{5}_{n_{0},n_{2}}$.

A reference point is
$
[0,{\bf 0};n_{0},n_{2}]
$
and we have:
$$
H^{F}_{[0,{\bf 0};n_{0},n_{2}]} = \{(0,{\bf v};\zeta_{1}) \vert
{\bf v} \in \R^{2}, \zeta_{1} \in {\bf T}\}
$$
i.e. a central extension of the Abelian group
$\R^{2}.$
Let
$\pi$
be an unitary irreducible representation of this group. Because
of (3.12) we must have:
$$
\pi_{(0,{\bf 0};\zeta_{1})} = \zeta_{1}^{-1} \times id
\eqno(3.18)$$

This easily implies that we have:
$$
\pi_{(0,{\bf v};1)}\pi_{(0,{\bf v'};1)} = exp\left\{-\frac{iF}{2}<{\bf v},
{\bf v'}>\right\} \pi_{(0,{\bf v}+{\bf v'};1)}
\eqno(3.19)$$
i.e.
$
{\bf v} \mapsto \pi_{(0,{\bf v};1)}
$
is an unitary irreducible representation of the canonical
commutation relations in Weyl form. According to Stone-von
Neumann theorem there exists (up to unitary equivalence) exactly
one such representation, denoted
$\pi^{CCR}$
and acting in the Hibert space
${\cal K}$
(for an explicit expression see e.g. \cite{T}, ch. 3.1). So, the
representations of
$
H^{F}_{[0,{\bf 0};n_{0},n_{2}]}
$
we are looking for are of the form:
$$
\pi_{(0,{\bf v};\zeta_{1})} = \zeta_{1}^{-1} \pi^{CCR}_{\bf v}.
\eqno(3.20)$$

A corresponding cocycle is clearly:
$$
\phi^{CCR}((x,{\bf v};\zeta_{1}),[p_{0},{\bf 0};n_{0},n_{2}]) =
\zeta_{1}^{-1}~\pi^{CCR}_{\bf v}
\eqno(3.21)$$

If we identify naturally
$Z^{5} \simeq \R$
with the strictly invariant measure
$dp_{0}$,
then the corresponding induced representation is acting in
$
{\cal H} = L^{2}(\R,{\cal K},dp_{0})
$
as follows:
$$
\left(W_{(x,{\bf v},\eta,{\bf
a};\zeta_{0},\zeta_{1},\zeta_{2})}f\right) (p_{0}) =
\zeta_{0}^{n_{0}}~\left(\zeta_{2}e^{-iS\eta
x}\right)^{n_{2}}~e^{i\eta
p_{0}}~\zeta_{1}^{-1}~\left(\pi^{CCR}_{\bf v} f\right)(p_{0}-Sn_{2}x)
\eqno(3.22)$$

Again (3.12) imposes
$
n_{0} = n_{2} = -1
$
and we are left with the projective representations
$V^{CCR}$
acting in
$
{\cal H} = L^{2}(\R,{\cal K},dp_{0})
$
according to:
$$
\left(V^{CCR}_{x,{\bf v},\eta,{\bf a}}f\right)(p_{0}) =
e^{i\eta(p_{0}+Sx)}~\left(\pi_{\bf v}^{CCR}f\right)(p_{0}+Sx)
\eqno(3.23)$$

b2) $Z = Z^{6}_{n_{0},n_{2},\rho}$.

A reference point is
$
[0,r{\bf e_{1}};n_{0},n_{2}]
$
and one easily obtains:
$$
H^{F}_{[0,r{\bf e}_{1};n_{0},n_{2}]} = \left\{\left(2\pi
n,\frac{2\pi nSn_{2}}{r}{\bf e_{1}}+\alpha {\bf
e_{2}};\zeta_{1}\right) \vert n \in {\bf Z}, \alpha \in \R,
\zeta \in {\bf T}\right\}
$$

If we denote:
$$
(n,\alpha;\zeta_{1}) \equiv \left(2\pi n,\frac{2\pi nSn_{2}}{r}
{\bf e_{1}}+\alpha {\bf e_{2}};\zeta_{1}\right)
$$
then the composition law is:
$$
(n,\alpha;\zeta_{1})\cdot(n',\alpha';\zeta'_{1}) =
(n+n',\alpha+\alpha';\zeta_{1}\zeta'_{1}exp\{ik(\alpha
n'-\alpha' n)\})
\eqno(3.24)$$
i.e. the little group is in this case a central extension of the
Abelian group
$
{\bf Z} \times \R.
$
Here
$
k = \frac{2\pi FSn_{2}}{r}
$.
Let
$
(n,\alpha;\zeta_{1}) \mapsto \pi_{(n,\alpha;\zeta_{1})}
$
be an unitary irreducible representation fulfilling:
$$
\pi_{(0,0;\zeta_{1})} = \zeta_{1}^{-1} \times id
\eqno(3.25)$$
(The argument leading to this relation is similar to the
argument leading to (3.18)).

Then we have for
$
\pi_{n,\alpha} \equiv \pi_{n,\alpha;1}
$:
$$
\pi_{n,\alpha}~\pi_{n',\alpha'} = exp\{ik(\alpha n'-\alpha' n)\}~
\pi_{n+n',\alpha+\alpha'}
\eqno(3.26)$$

The resemblance of this relation to the Weyl system (3.19)
suggests us to associate to (3.26) a sort of imprimitivity
system \cite{V}. We denote:
$
\pi_{1,0} \equiv V
$
and
$
\pi_{\alpha} = \pi_{0,\alpha}
$.
It is clear that (3.26) is equivalent to:
$$
\pi_{n,0} = V^{n};~~V\pi_{\alpha}V^{-1} =
e^{2ik\alpha}\pi_{\alpha};~~\pi_{\alpha} \pi_{\alpha'} =
\pi_{\alpha+\alpha'}
\eqno(3.27)$$

So, it is sufficient to find an irreducible couple
$
(V,\pi_{\alpha})
$
fulfilling the last two relations of (3.27). According to GNS
theorem
$
\pi_{\alpha}
$
is of the form:
$$
\pi_{\alpha} = \int_{\R} e^{i\lambda\alpha} dP(\lambda)
\eqno(3.28)$$
where
$
\Delta \mapsto P_{\Delta}
$
is a projection valued measure in the Hilbert space
${\cal K}$.
The second relation (3.27) is equivalent to:
$$
V P_{\Delta} V^{-1} = P_{\Delta-2k}
\eqno(3.29)$$

Applying the same idea as in \cite{V} Lemma 6.10, we can prove that
$P$
is homogeneous i.e.
$
{\cal K} = L^{2}(\R,{\bf C}^{n},\beta)
$
where
$\beta$
is measure on
$\R$
quasi-invariant with respect to the transformation
$
\lambda \mapsto \lambda+2k
$
and
$$
(P_{\Delta}f)(\lambda) = \chi_{\Delta}(\lambda)f(\lambda)
$$.

Moreover, like in \cite{V} Thm. 6.12 one can show that
$V$
has the "diagonal" expression:
$$
(Vf)(\lambda) = r_{V}(\lambda+2k)^{1/2} v(\lambda) f(\lambda+2k)
$$
(here
$r_{V}$
is a Radon-Nycodim derivative).

It is easy to see that for
$n > 1$
the system
$
(V,P_{\Delta})
$
is not irreducible. So,
$n = 1$.
Moreover one can show that
$
supp(\beta)
$
must be discrete with period
$2k$
i.e.
$
supp(\beta) = \{\lambda+2km \vert m \in {\bf
Z}\},~\lambda \in \R.
$

It is clear that we can take
$
\beta = \sum_{m \in {\bf Z}} \delta_{\lambda+2km}.
$
In this case one can also take
${\cal K} = l^{2},
$
(i. e. sequences
$
\{f_{m}\}_{m \in {\bf Z}}
$
such that
$
\sum_{m \in {\bf Z}} \vert f_{m}\vert^{2} < \infty
$)
and
$
P = \beta.
$
It follows that we have:
$$
\left(\pi_{\alpha}f\right)_{m} = e^{i(\lambda+2km)\alpha} f_{m};~
(Vf)_{m} = v_{m} f_{m+1}~(\forall m \in {\bf Z})
\eqno(3.30)$$
(Here
$v_{m} \in {\bf T}$).

The system
$
(V,\pi_{\alpha})
$
is irreducible. So, the unitary irreducible representations of
the little group
$
H^{F}_{[0,r{\bf e_{1}};n_{0},n_{2}]}
$
verifying (3.25) are of the form
$
\pi^{\lambda}~(\lambda \in \R)
$
acting in
$
l^{2}
$
according to:
$$
\left(\pi^{\lambda}_{n,\alpha;\zeta_{1}}f\right)_{m} =
\zeta_{1}^{-1} e^{i[\lambda+k(2m+n)]\alpha} f_{m+n}
\eqno(3.31)$$

We denote by
$\phi^{\lambda}
$
a cocycle corresponding to
$
\pi^{\lambda}
$.
(We have not been able to find a simple expression for this
cocycle). Then, if we identify
$
Z^{6} \simeq \R \times S^{1}
$
with the strictly invariant measure
$dp_{0}d\Omega
$,
we get in the end the projective representation
$
V^{r,\lambda}
$
acting in
$L^{2}(\R \times S^{1},l^{2},dp_{0}d\Omega)
$
according to:
$$
\left( V^{r,\lambda}_{x,{\bf v},\eta,{\bf a}}
f\right)(p_{0},{\bf p}) = exp\{i[\eta(p_{0}+Sx)+{\bf a}\cdot
{\bf p}]\} \times$$
$$\phi^{\lambda}((x,{\bf v};1),[p_{0}+{\bf v}\cdot
{\bf p}+Sx,R(x)^{-1}{\bf p}]) f(p_{0}+{\bf v}\cdot {\bf p}+
Sx,R(x)^{-1}{\bf p})
\eqno(3.32)$$

c) $\tau \not= 0,~S \not= 0$

Only the orbit
$
Z^{10}_{n_{0},n_{2}}
$
is involved. A reference point is
$
[0,{\bf 0};n_{0},n_{2}]
$
and we have:
$$
H^{F}_{[0,{\bf 0};n_{0},n_{2}]} = \{(0,{\bf 0};\zeta_{1}) \vert
\zeta_{1} \in {\bf T} \} \simeq {\bf T}
$$
with the unitary irreducible representations of the form
$
\pi^{n_{1}}
$
acting in
${\bf C}$
according to:
$$
\pi^{n_{1}}_{(0,{\bf 0};\zeta_{1})} = \zeta_{1}^{n_{1}}
\eqno(3.33)$$

A corresponding cocycle can be found:
$$
\phi^{n_{1}}((x,{\bf v};\zeta_{1}),[p_{0},{\bf p};n_{0},n_{2}])
= \zeta_{1}^{n_{1}} exp\left\{ \frac{iFn_{1}}{2n_{0}\tau} <{\bf
v},R(x){\bf p}>\right\}
\eqno(3.34)$$

We identify
$
Z^{10} \simeq \R \times \R^{2}
$
with the strictly invariant measure
$
dp_{0} d{\bf p}
$.
The induced representations obey (3.12) {\it iff}
$
n_{0} = n_{1} =n_{2} = -1
$
and we get projective representations
$V^{\tau}$
acting in
$L^{2}(\R \times \R^{2},dp_{0} d{\bf p})
$
acording to:
$$
\left(V^{\tau}_{x,{\bf v},\eta,{\bf a}} f\right)(p_{0},{\bf p})
= exp\left\{i\left[ \eta(p_{0}+Sx)+\frac{\tau}{2} {\bf a}\cdot {\bf v}+
{\bf a}\cdot {\bf p}+\frac{F}{2\tau}<{\bf v},{\bf p}>\right]\right\}$$
$$f(p_{0}+Sx+{\bf v}\cdot {\bf
p}+\frac{1}{2}\tau{\bf v}^{2},R(x)^{-1}({\bf p}+\tau{\bf v})).
\eqno(3.35)$$

So every projective unitary irreducible representation of
$G^{\uparrow}_{+}$
with a non-trivial multiplicator is unitary equivalent to a
representation of the type
$
V^{\tau,s},~V^{CCR},~V^{r,\lambda}
$,
or
$V^{\tau}$
described by (3.17), (3.23), (3.32) and (3.35) respectively.

\subsection*{D. Infinitesimal generators}

The representations (3.17) obtained in case a) are the analogues
of the representations obtained in 1+3 dimensions (see \cite{V},
ch. IX, subsec. 8). The infinitesimal generators of this
representation are:
$$
(H f)({\bf p}) = \frac{{\bf p}^{2}}{2\tau} f
\eqno(3.36)$$
$$
({\bf P} f)({\bf p}) = {\bf p} f
\eqno(3.37)$$
$$
(J f)({\bf p}) = i\left( p_{1} {\partial f\over \partial p_{2}}
- p_{2} {\partial f\over \partial p_{1}}\right) + sf
\eqno(3.38)$$
$$
({\bf K} f)({\bf p}) = -i\tau{\partial f\over \partial {\bf p}} +
\frac{F}{2\tau} A{\bf p} f
\eqno(3.39)$$

So
$\tau$
and
$s$
must be interpreted as the mass and the spin of the system
respectively. The interpretation of $F$
is not clear. Because
$F$
appears from a central extension of a translation group (see
(3.19)), it is tempting to associate
$F$
with some kind of magnetic force. By restricting to the covering
group of the Euclidean group
$SE(2)$
we obtain:
$$
\left( V^{\tau,s}_{x,{\bf a}} f\right)({\bf p}) = e^{i(sx+
{\bf a}\cdot {\bf p})} f(R(x)^{-1}{\bf p})
\eqno(3.40)$$
so performing a Fourier transform we can conclude that the system
is localisable on
$\R^{2}$:
$$
\left( V^{\tau,s}_{x,{\bf a}} f\right)({\bf X}) = e^{isx}
f(R(x)^{-1} ({\bf X} - {\bf a}))
\eqno(3.41)$$

The representations (3.23) and (3.32) from case b) are not
localisable on
$\R^{2}$
(the argument is similar to the one in \cite{V} and is based on
the existence of the constrain
$
{\bf p}^{2} = r^{2}
$).

In case c) the infinitesimal generators are:
$$
H f = p_{0} f
\eqno(3.42)$$
$$
{\bf P} f = {\bf p} f
\eqno(3.43)$$
$$
J f = i\left( p_{1} {\partial f\over \partial p_{2}}
- p_{2} {\partial f\over \partial p_{1}} - S {\partial
f\over \partial p_{0}}\right)
\eqno(3.44)$$
$$
{\bf K} f = -i\left( \tau{\partial f\over \partial {\bf p}} +
{\bf p} {\partial f\over \partial p_{0}}\right) +
\frac{F}{2\tau} A{\bf p} f
\eqno(3.45)$$

So in this case
$S$
can be interpreted as the spin of the system. For
$\tau$
and
$F$
it is natural to keep the interpretations from case a).

By restricting to
$
\widetilde{SE(2)}
$
we obtain:
$$
\left( V^{\tau}_{x,{\bf a}} f\right)(p_{0},{\bf p}) = e^{i{\bf
a}\cdot {\bf p}} f(p_{0}+Sx,R(x)^{-1}{\bf p}))
\eqno(3.46)$$

By performing a three-dimensional Fourier transform we get:
$$
\left( V^{\tau}_{x,{\bf a}} f\right)(X_{0},{\bf X}) =
e^{iSX_{0}x} f(X_{0},R(x)^{-1}({\bf X}-{\bf a}))
\eqno(3.47)$$

The formula shows that the system is localisable on
$\R^{2}$.

The appearance of the representation
$V^{\tau}$
is rather intriguing. Indeed in the relativistic case \cite{G},
there exists a single class of projective representation of the
Poincar\'e group corresponding to non-zero mass systems and one
would expect that in the non-relativistic limit we obtain only
the representations from case a) (like in 1+3 dimensions).

Finally, we note that the true representations of
$
G^{\uparrow}_{+}
$
are easy to find (see \cite{G} Thm. 2) but they do not
correspond to localisable systems.


\begin{thebibliography}{9}

\bibitem{G}
D. R. Grigore, "The Projective Unitary Irreducible
Representations of the Poincar\'e Group in 1+2 Dimensions",
Journ. Math. Phys. 34 (1993) 4172-4189

\bibitem{MS}
J. Mund, R. Schrader, "Hilbert Space for Non-relativistic and
Relativistic ``Free'' Plektons (Particles with Braid Group
Statistics)", preprint Univ. Berlin, SFB 288 no. 74

\bibitem{V}
V. S. Varadarajan, ``Geometry of Quantum Theory'' (second edition),
Springer, New York, 1985

\bibitem{T}
W. Thirring, "Quantum Mechanics of Atoms and Molecules",
Springer, New York, 1979

\end{thebibliography}
\end{document}